\pdfoutput=1

\documentclass[12pt,a4paper]{article}

\usepackage{ifthen} 
\newboolean{pdflatex}
\setboolean{pdflatex}{true} 

\newboolean{articletitles}
\setboolean{articletitles}{true} 

\newboolean{uprightparticles}
\setboolean{uprightparticles}{false} 

\newboolean{inbibliography}
\setboolean{inbibliography}{false} 

\def\paperauthors{LHCb collaboration} 
\def\paperasciititle{JPsi UPC} 
\def\papertitle{Study of coherent \jpsi production \\ in lead-lead collisions \\ at $\sqsnn=5\tev$ 
} 
\def\paperkeywords{{High Energy Physics}, {LHCb}} 
\def\papercopyright{\the\year\ CERN for the benefit of the LHCb collaboration} 
\def\paperlicence{CC BY 4.0 licence}
\def\paperlicenceurl{https://creativecommons.org/licenses/by/4.0/}
\def\update{{\color{red}}}
\def\lumimeanrounded {\ensuremath{10}}
\def\lumiubarn {\ensuremath{10.1\pm1.3}}

\def\coherentcrosssectionuncsplit {\ensuremath{4.45 \pm {0.24} \pm {0.18} \pm {0.58}\update}}

\usepackage[top=1in, bottom=1.25in, left=1in, right=1in]{geometry}

\usepackage{microtype}
\usepackage{lineno}  
\usepackage{xspace} 
\usepackage{caption} 

\usepackage{graphicx}  
\usepackage{color}
\usepackage{colortbl}
\graphicspath{{./figs/}} 

\usepackage{amsmath} 
\usepackage{amssymb}
\usepackage{amsfonts}
\usepackage{upgreek} 

\newcommand*\patchAmsMathEnvironmentForLineno[1]{%
\expandafter\let\csname old#1\expandafter\endcsname\csname #1\endcsname
\expandafter\let\csname oldend#1\expandafter\endcsname\csname
end#1\endcsname
 \renewenvironment{#1}%
   {\linenomath\csname old#1\endcsname}%
   {\csname oldend#1\endcsname\endlinenomath}%
}
\newcommand*\patchBothAmsMathEnvironmentsForLineno[1]{%
  \patchAmsMathEnvironmentForLineno{#1}%
  \patchAmsMathEnvironmentForLineno{#1*}%
}
\AtBeginDocument{%
\patchBothAmsMathEnvironmentsForLineno{equation}%
\patchBothAmsMathEnvironmentsForLineno{align}%
\patchBothAmsMathEnvironmentsForLineno{flalign}%
\patchBothAmsMathEnvironmentsForLineno{alignat}%
\patchBothAmsMathEnvironmentsForLineno{gather}%
\patchBothAmsMathEnvironmentsForLineno{multline}%
\patchBothAmsMathEnvironmentsForLineno{eqnarray}%
}

\usepackage{hyperxmp}

\usepackage[pdftex,
            pdfauthor={\paperauthors},
            pdftitle={\paperasciititle},
            pdfkeywords={\paperkeywords},
            pdfcopyright={Copyright (C) \papercopyright},
            pdflicenseurl={\paperlicenceurl}]{hyperref}

\usepackage[all]{hypcap} 

\usepackage{xspace} 
\usepackage{upgreek}

\def\lhcb   {\mbox{LHCb}\xspace}

\def\MagUp {\mbox{\em Mag\kern -0.05em Up}\xspace}

\ifthenelse{\boolean{uprightparticles}}%
{
 
 \def\Pgamma      {\ensuremath{\upgamma}\xspace}

 \def\Pmu         {\ensuremath{\upmu}\xspace}

 \def\Ppi         {\ensuremath{\uppi}\xspace}

 \def\Ppsi        {\ensuremath{\uppsi}\xspace}

 \def\PDelta      {\ensuremath{\Delta}\xspace}                 
 \def\PXi      {\ensuremath{\Xi}\xspace}                 
 \def\PLambda      {\ensuremath{\Lambda}\xspace}                 
 \def\PSigma      {\ensuremath{\Sigma}\xspace}                 
 \def\POmega      {\ensuremath{\Omega}\xspace}                 
 \def\PUpsilon      {\ensuremath{\Upsilon}\xspace}                 
 
 \def\PB      {\ensuremath{\mathrm{B}}\xspace}                 
                  
 \def\PD      {\ensuremath{\mathrm{D}}\xspace}

 \def\PJ      {\ensuremath{\mathrm{J}}\xspace}                 
 \def\PK      {\ensuremath{\mathrm{K}}\xspace}

 \def\PX      {\ensuremath{\mathrm{X}}\xspace}

 \def\Pb      {\ensuremath{\mathrm{b}}\xspace}                 
 \def\Pc      {\ensuremath{\mathrm{c}}\xspace}

 \def\Pi      {\ensuremath{\mathrm{i}}\xspace}

}
{
 
 \def\Pgamma      {\ensuremath{\gamma}\xspace}

 \def\Pmu         {\ensuremath{\mu}\xspace}

 \def\Ppi         {\ensuremath{\pi}\xspace}

 \def\Ppsi        {\ensuremath{\psi}\xspace}                 
                  
 \mathchardef\PDelta="7101
 \mathchardef\PXi="7104
 \mathchardef\PLambda="7103
 \mathchardef\PSigma="7106
 \mathchardef\POmega="710A
 \mathchardef\PUpsilon="7107
                  
 \def\PB      {\ensuremath{B}\xspace}                 
                  
 \def\PD      {\ensuremath{D}\xspace}

 \def\PJ      {\ensuremath{J}\xspace}                 
 \def\PK      {\ensuremath{K}\xspace}

 \def\PX      {\ensuremath{X}\xspace}

 \def\Pb      {\ensuremath{b}\xspace}                 
 \def\Pc      {\ensuremath{c}\xspace}

 \def\Pi      {\ensuremath{i}\xspace}

}

\makeatletter
\ifcase \@ptsize \relax
  \newcommand{\miniscule}{\@setfontsize\miniscule{4}{5}}
\or
  \newcommand{\miniscule}{\@setfontsize\miniscule{5}{6}}
\or
  \newcommand{\miniscule}{\@setfontsize\miniscule{5}{6}}
\fi
\makeatother

\DeclareRobustCommand{\optbar}[1]{\shortstack{{\miniscule (\rule[.5ex]{1.25em}{.18mm})}
  \\ [-.7ex] $#1$}}

\def\mup        {{\ensuremath{\Pmu^+}}\xspace}
\def\mun        {{\ensuremath{\Pmu^-}}\xspace} 

\def\mumu       {{\ensuremath{\Pmu^+\Pmu^-}}\xspace}

\def\cquark    {{\ensuremath{\Pc}}\xspace}

\def\bquark    {{\ensuremath{\Pb}}\xspace}

\def\pion   {{\ensuremath{\Ppi}}\xspace}

\def\pip    {{\ensuremath{\pion^+}}\xspace}
\def\pim    {{\ensuremath{\pion^-}}\xspace}

  \def\Kbar    {{\kern 0.2em\overline{\kern -0.2em \PK}{}}\xspace}

\def\KorKbar    {\kern 0.18em\optbar{\kern -0.18em K}{}\xspace}

  \def\Dbar    {{\kern 0.2em\overline{\kern -0.2em \PD}{}}\xspace}

\def\DorDbar    {\kern 0.18em\optbar{\kern -0.18em D}{}\xspace}

\def\Bbar    {{\ensuremath{\kern 0.18em\overline{\kern -0.18em \PB}{}}}\xspace}

\def\BorBbar    {\kern 0.18em\optbar{\kern -0.18em B}{}\xspace}

\def\jpsi     {{\ensuremath{{\PJ\mskip -3mu/\mskip -2mu\Ppsi\mskip 2mu}}}\xspace}
\def\psitwos  {{\ensuremath{\Ppsi{(2S)}}}\xspace}

  \def\Y#1S{\ensuremath{\PUpsilon{(#1S)}}\xspace}

\def\Lbar        {{\ensuremath{\kern 0.1em\overline{\kern -0.1em\PLambda}}}\xspace}
\def\LorLbar    {\kern 0.18em\optbar{\kern -0.18em \PLambda}{}\xspace}

\def\BF         {{\ensuremath{\mathcal{B}}}\xspace}

\def\BR         {\BF}
\newcommand{\decay}[2]{\ensuremath{#1\!\to #2}\xspace}         

\def\to                 {\ensuremath{\rightarrow}\xspace}

\def\AT#1     {\ensuremath{A_{\mathrm{T}}^{#1}}\xspace}           

\def\C#1      {\ensuremath{\mathcal{C}_{#1}}\xspace}                       
\def\Cp#1     {\ensuremath{\mathcal{C}_{#1}^{'}}\xspace}                    
\def\Ceff#1   {\ensuremath{\mathcal{C}_{#1}^{\mathrm{(eff)}}}\xspace}        
\def\Cpeff#1  {\ensuremath{\mathcal{C}_{#1}^{'\mathrm{(eff)}}}\xspace}       
\def\Ope#1    {\ensuremath{\mathcal{O}_{#1}}\xspace}                       
\def\Opep#1   {\ensuremath{\mathcal{O}_{#1}^{'}}\xspace}                    

\newcommand{\tev}{\ifthenelse{\boolean{inbibliography}}{\ensuremath{~T\kern -0.05em eV}}{\ensuremath{\mathrm{\,Te\kern -0.1em V}}}\xspace}
\newcommand{\gev}{\ensuremath{\mathrm{\,Ge\kern -0.1em V}}\xspace}
\newcommand{\mev}{\ensuremath{\mathrm{\,Me\kern -0.1em V}}\xspace}
\newcommand{\kev}{\ensuremath{\mathrm{\,ke\kern -0.1em V}}\xspace}
\newcommand{\ev}{\ensuremath{\mathrm{\,e\kern -0.1em V}}\xspace}
\newcommand{\gevc}{\ensuremath{{\mathrm{\,Ge\kern -0.1em V\!/}c}}\xspace}
\newcommand{\mevc}{\ensuremath{{\mathrm{\,Me\kern -0.1em V\!/}c}}\xspace}
\newcommand{\gevcc}{\ensuremath{{\mathrm{\,Ge\kern -0.1em V\!/}c^2}}\xspace}
\newcommand{\gevgevcccc}{\ensuremath{{\mathrm{\,Ge\kern -0.1em V^2\!/}c^4}}\xspace}
\newcommand{\mevcc}{\ensuremath{{\mathrm{\,Me\kern -0.1em V\!/}c^2}}\xspace}

\def\m    {\ensuremath{\mathrm{ \,m}}\xspace}

\def\mbarn{\ensuremath{\mathrm{ \,mb}}\xspace}

\def\invmub{\ensuremath{{\mathrm{ \,\upmu b}^{-1}}}\xspace}

\def\gsim{{~\raise.15em\hbox{$>$}\kern-.85em
          \lower.35em\hbox{$\sim$}~}\xspace}
\def\lsim{{~\raise.15em\hbox{$<$}\kern-.85em
          \lower.35em\hbox{$\sim$}~}\xspace}

\def\sqs   {\ensuremath{\protect\sqrt{s}}\xspace}
\def\sqsnn {\ensuremath{\protect\sqrt{s_{\scriptscriptstyle\rm NN}}}\xspace}
\def\ptjpsisq   {\ensuremath{p_{\mathrm{ T}}^{2}}\xspace}
\def\ptjpsi     {\ensuremath{p_{\mathrm{ T}}}\xspace}
\def\yjpsi      {\ensuremath{{y}}\xspace}

\def\pt         {\ensuremath{p_{\mathrm{ T}}}\xspace}
\def\ptot       {\ensuremath{p}\xspace}

\def\tell1  {TELL1\xspace}
\def\ukl1   {UKL1\xspace}

\usepackage{cite} 
\usepackage{mciteplus}

\usepackage{longtable} 
\usepackage[bottom]{footmisc}
\begin{document}

\renewcommand{\thefootnote}{\fnsymbol{footnote}}
\setcounter{footnote}{1}


\begin{titlepage}
\pagenumbering{roman}

\vspace*{-1.5cm}
\centerline{\large EUROPEAN ORGANIZATION FOR NUCLEAR RESEARCH (CERN)}
\vspace*{1.5cm}
\noindent
\begin{tabular*}{\linewidth}{lc@{\extracolsep{\fill}}r@{\extracolsep{0pt}}}
\ifthenelse{\boolean{pdflatex}}
{\vspace*{-1.5cm}\mbox{\!\!\!\includegraphics[width=.14\textwidth]{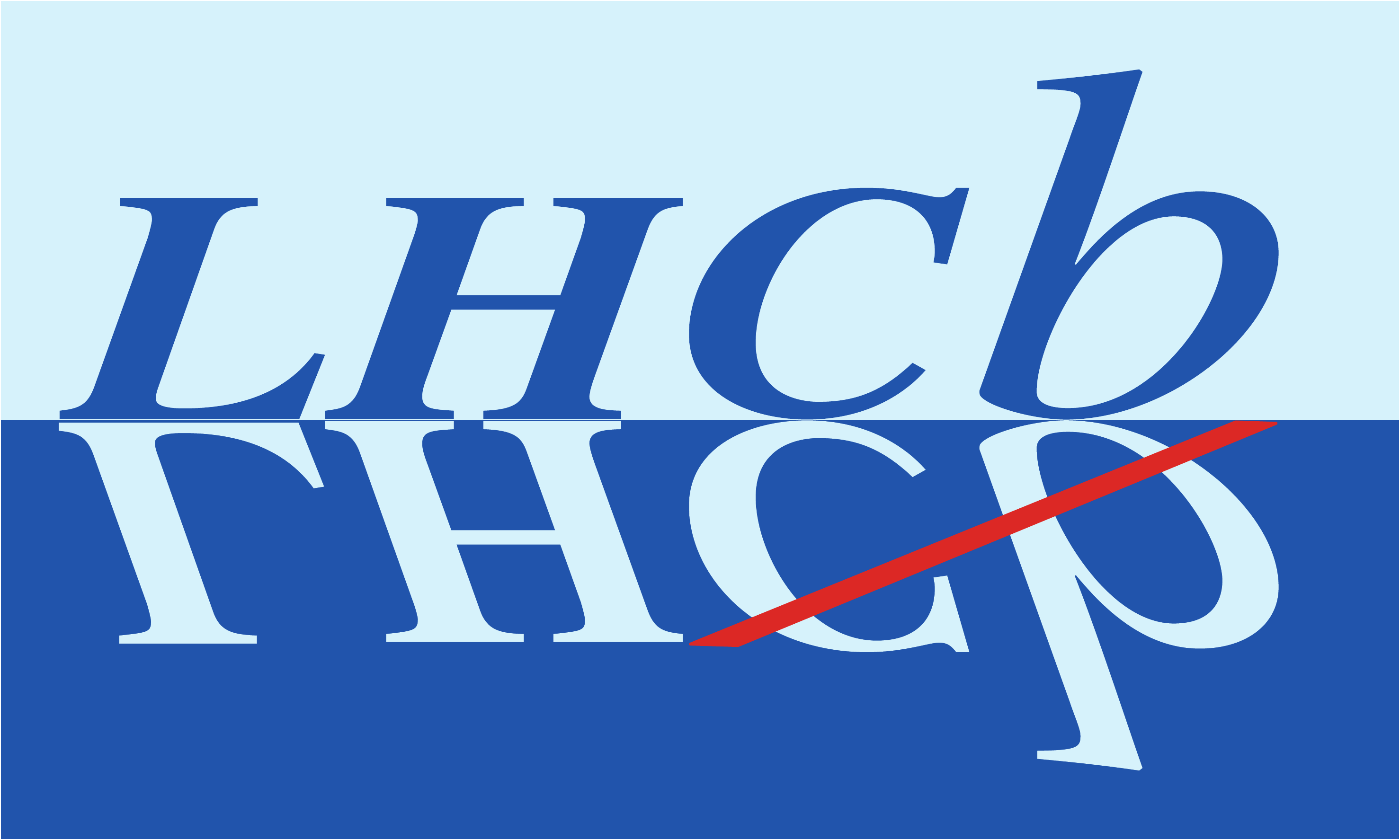}} & &}%
{\vspace*{-1.2cm}\mbox{\!\!\!\includegraphics[width=.12\textwidth]{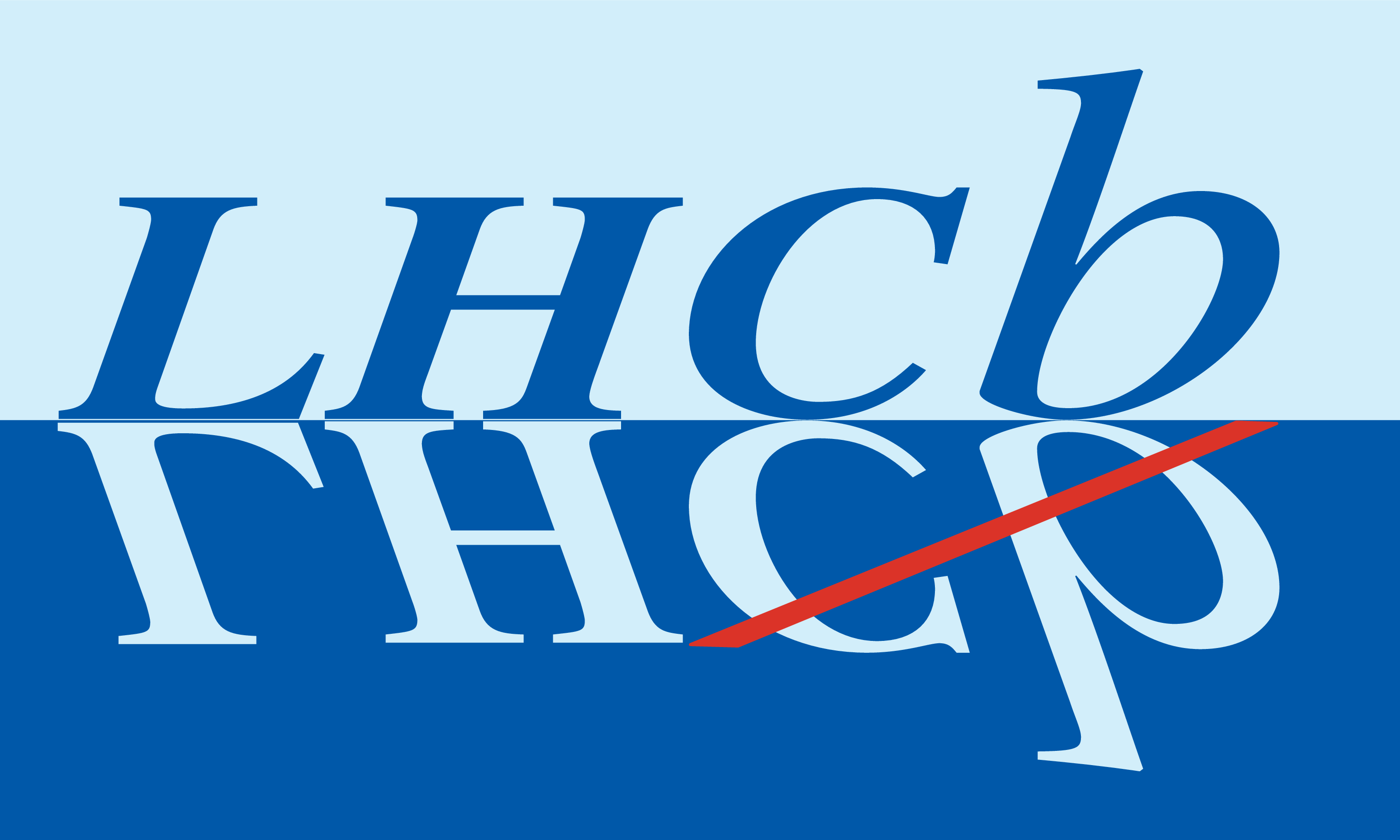}} & &}%
\\
 & & CERN-EP-2021-108 \\  
 & & LHCb-PAPER-2021-013 \\  
 & & 19 July 2022 \\ 
\end{tabular*}

\vspace*{4.0cm}

{\normalfont\bfseries\boldmath\huge
\begin{center}
  \papertitle 
\end{center}
}

\vspace*{2.0cm}

\begin{center}
\paperauthors\footnote{Authors are listed at the end of this paper.}
\end{center}

\vspace{\fill}

\begin{abstract}
  \noindent
Coherent production of \jpsi mesons is studied in ultraperipheral lead-lead collisions at a nucleon-nucleon centre-of-mass energy of $5\tev$, using a data sample collected by the LHCb experiment corresponding to an integrated luminosity of about $\lumimeanrounded\invmub$.
The \jpsi mesons are reconstructed in the dimuon final state and are required to have transverse momentum below $1\gev$.
The cross-section within the rapidity range of $2.0 < y < 4.5$ is measured to be \mbox{$\coherentcrosssectionuncsplit\mbarn$}, where the first uncertainty is statistical, the second systematic and the third originates from the luminosity determination.
The cross-section is also measured in \jpsi rapidity intervals.
The results are compared to predictions from phenomenological models.  
\end{abstract}

\vspace*{1.5cm}

\begin{center}
  Published in JHEP 07 (2022) 117
\end{center}

\vspace{\fill}

{\footnotesize 
\centerline{\copyright~\papercopyright. \href{\paperlicenceurl}{\paperlicence}.}}
\vspace*{2mm}

\end{titlepage}


\newpage
\setcounter{page}{2}
\mbox{~}
%
%
%
%

\cleardoublepage


\renewcommand{\thefootnote}{\arabic{footnote}}
\setcounter{footnote}{0}



\pagestyle{plain} 
\setcounter{page}{1}
\pagenumbering{arabic}


%


\section{Introduction}
\label{sec:Introduction}

In ultra-relativistic collisions of heavy-nuclei at the LHC, vector mesons can be produced through
two-photon and photonuclear interactions in ultraperipheral collisions (UPCs), where 
the two nuclei collide with an impact parameter larger than the sum of their radii.
The cross-sections for photon-induced reactions are large because the intensity of the photon flux is enhanced by the strong electromagnetic field of the nucleus, which increases with the square of the atomic number. %
The interactions are either coherent, where the photon couples to all nucleons, or incoherent, where the photon couples to a single nucleon. 
In the incoherent case the nucleus is likely to break up, leading to a higher transverse momentum, \ptjpsi, of the meson.
Coherent \jpsi-meson production in UPCs can be described by the interaction of photons with gluons identified as a single object with vacuum quantum numbers.
This is commonly modelled as a pomeron (I$\!$P) exchange~\cite{Guzey:2016piu,starlight,PhysRevC.85.044904,PhysRevC.84.011902,PhysRevC.86.014905}. 

\begin{figure}[b] 
  \begin{center}
    \includegraphics[width=0.35\linewidth]{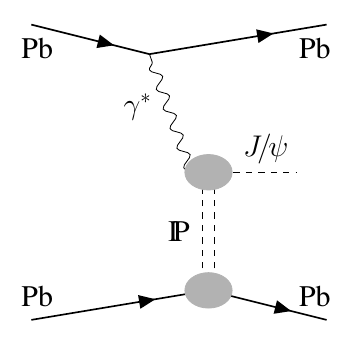}
    \hspace{1cm}
    \includegraphics[width=0.35\linewidth]{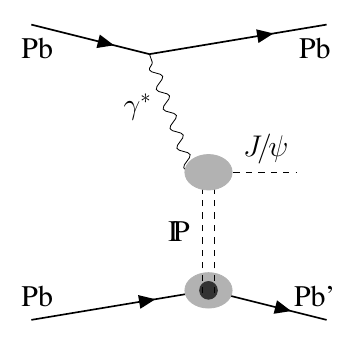}
  \end{center}
  \caption{Illustration of the (left) coherent scatter with the lead nucleus and (right) incoherent interaction with a single nucleon leading to exclusive production of \jpsi mesons in ultraperipheral heavy-ion collisions. The symbol Pb' represents any final state for the nucleus inelastic scattering in the incoherent process.}
  \label{fig:feynman}
\end{figure}

An illustration of this process is given in Fig.~\ref{fig:feynman}.
This interaction probes the gluon distribution at a hard momentum transfer $Q^2$ of about $m_\jpsi^2/4$, where $m_\jpsi$ is the \jpsi mass~\cite{Bertulani:2005ru,Ryskin}.%
\footnote{In this paper natural units where $c$ = 1 are used.} 
%
%
In this paper, a measurement of coherent \jpsi production is reported in lead-lead collisions at a nucleon-nucleon centre-of-mass energy of $\sqsnn=5\tev$ collected with the LHCb detector in 2015, corresponding to an integrated luminosity of about $\lumimeanrounded\invmub$.
The forward rapidity range $2.0<\yjpsi<4.5$ covered by the present 
measurement corresponds to values of the Bjorken variable $x\approx(m_\jpsi/\sqsnn)e^{\pm \yjpsi}$ down to $10^{-5}$.
At these $x$ values, current uncertainties on the gluon distributions inside the nucleon are sizeable~\cite{Eskola:2009uj,Eskola:2016oht}, thus new measurements 
should reduce the \mbox{uncertainties~\cite{Flett:2019nga,Flett:2019pux,Kovarik:2015cma}}.
%
Results of UPC studies have also been reported by RHIC and LHC experiments~\cite{Afanasiev:2009hy,Adams:2004rz,Khachatryan:2016qhq,Abelev:2012ba,Abbas:2013oua,TheALICE:2014dwa,Adam:2015gsa,Acharya:2019vlb,ALICE:2021jnv,ALICE:2021gpt,ALICE:2021tyx}.
In particular, recent measurements in the forward rapidity region $2.5<\yjpsi<4.0$~\cite{Acharya:2019vlb} by the ALICE Collaboration show the importance of including nuclear modification effects in theoretical models.

The paper is organised as follows. 
The LHCb detector and the event selection are described in Sec.~\ref{sec:detector}. The analysis strategy and the systematic uncertainties are discussed in Secs.~\ref{sec:strategy} and \ref{sec:sys}, respectively. The differential cross-section results for \jpsi production in UPCs are detailed in Sec.~\ref{sec:results}. Conclusions are given in Sec.~\ref{sec:conclusion}.

\section{Detector description and candidate selection}
\label{sec:detector}

The \lhcb detector~\cite{Alves:2008zz,LHCb-DP-2014-002} is a
single-arm forward spectrometer covering the \mbox{pseudorapidity}
range $2<\eta <5$, designed for the study of particles containing
\bquark or \cquark quarks. The detector includes a high-precision
tracking system consisting of a silicon-strip vertex detector (VELO)
surrounding the interaction region, a silicon-strip detector located
upstream of a dipole magnet with a bending power of about
$4{\mathrm{\,Tm}}$, and three stations of silicon-strip detectors and
straw drift tubes placed downstream of the magnet. The tracking system
provides a measurement of momentum, \ptot, of charged particles with a
relative uncertainty that varies from 0.5\% at low momentum to 1.0\%
at 200\gev.
The VELO has a material budget of one fifth of a radiation length that allows reconstruction and rejection of events with additional low-momentum tracks~\cite{LHCb-DP-2014-001}.
Different types of charged hadrons are distinguished using information
from two ring-imaging Cherenkov detectors. Photons, electrons and
hadrons are identified by a calorimeter system consisting of
scintillating-pad (SPD) and preshower (PRS) detectors, an
electromagnetic calorimeter and a hadronic calorimeter. Muons are
identified by a system composed of alternating layers of iron and
multiwire proportional chambers.  
The pseudorapidity coverage is extended by forward shower counters 
(HeRSCheL) consisting of five
planes of scintillators with three planes at $114$, $19.7$ and $7.5\m$ 
upstream of the interaction
point, and two downstream at $20$ and $114\m$. 
The HeRSCheL detector significantly extends the acceptance in which hadron showers can be detected 
to classify central exclusive production and UPC events by covering a pseudorapidity region of 
approximately $-10<\eta<-5$ and $5<\eta<10$~\cite{LHCb-DP-2016-003}.
The real-time event selection is performed by a trigger, which
consists of a hardware stage, based on information from the
calorimeter and muon systems, followed by a software stage, which
applies a full event reconstruction.

In this analysis, \jpsi candidates are selected through their decays into
two oppositely charged muons.
The events are
selected by the trigger system, requiring information from the
muon system to be compatible with at least one muon with
$\pt$ larger than $900\mev$ at the hardware level, and the
invariant mass of the two muons, $m_{\mu^+\mu^-}$,  exceeding $2.7\gev$ at the software level.
In the offline selection, candidates are identified by requiring both
muons to have $\pt>800\mev$ within the pseudorapidity region
$2.0<\eta<4.5$, and 
$m_{\mu^+\mu^-}$ to be within $65\mev$ of
the known \jpsi mass~\cite{PDG2020}. 
Only \jpsi candidates with reconstructed $\pt<1\gev$ and an azimuthal opening angle between the muons larger than $0.9\pi$ are retained.

In order to suppress background
from more central lead-lead collisions, events with more than 20 deposits in the SPD are vetoed.
In addition, events with an extra VELO track in the spatial vicinity of the reconstructed \jpsi candidate are rejected.
Finally, a requirement on the activity in the HeRSCheL detector,
based upon a figure-of-merit that combines
detector signals of all stations~\cite{LHCb-DP-2016-003}, is used to discard events with significant activity in the HeRSCheL acceptance region.

\section{Cross-section measurement}
\label{sec:strategy}

The differential cross-section for coherent \jpsi production is evaluated as
\begin{align}
  \frac{d\sigma
  }{dy}&=
\frac{n_{\text {coh}}}{\varepsilon_y \,\Delta y \,\mathcal{L} \,\BR} \,,
\label{eqn:dsigmady}
\end{align}
where $n_{\text {coh}}$ is the signal yield, $\varepsilon_y$ is the total
efficiency in each rapidity interval, $\Delta y$ is the rapidity interval width,
$\mathcal{L}$ is the integrated luminosity, and $\BR=(5.961\pm0.033)\% $ is the $\decay{\jpsi}{\mumu}$ branching
fraction \cite{PDG2020}.

The luminosity is determined with the same method as for proton-proton \mbox{collisions~\cite{LHCb-PAPER-2011-015,LHCb-PAPER-2014-047}}.
Interaction rates are determined from random luminosity triggers and are encapsulated in a small number of luminosity sensitive observables. These observables include the number of vertices and tracks in the VELO, the number of identified muons, the number of hits in the SPD,  and the transverse energy deposition in the calorimeters.
Whereas the rates from the different observables are consistent in proton-proton and proton-lead collisions, in lead-lead collisions some discrepancies are observed, which are used to assign a systematic uncertainty.
The integrated luminosity of the data sample is determined to be 
\lumiubarn\invmub,
where the absolute calibration is performed with Van der Meer scans~\cite{LHCb-PAPER-2014-047}.
The luminosity determination is checked using the ratio of \decay{\Pgamma\Pgamma}{\mumu} events over coherent \jpsi events in the range of the measurement. 
This ratio was found within $10\%$ of the ratio predicted by {\sc SuperChic 4}\cite{Harland-Lang:2020veo} using the {\tt MMHT2015qed NNLO}  probability density function setting\cite{Harland-Lang:2019pla} for both the differential and the integrated cross-section ratio.


\subsection{Signal yield determination}

The signal yield is determined in two steps. First,
a fit to the dimuon invariant mass spectrum is performed to obtain the
\jpsi yield, which includes the contribution of coherent and incoherent
\jpsi mesons, and feed-down from \jpsi mesons originating from \psitwos decays. 
Second, a fit to the \jpsi
transverse momentum
is used to isolate the coherent \jpsi yield. 

\begin{figure}[htb]
  \begin{center}
    \includegraphics[width=0.98\linewidth]{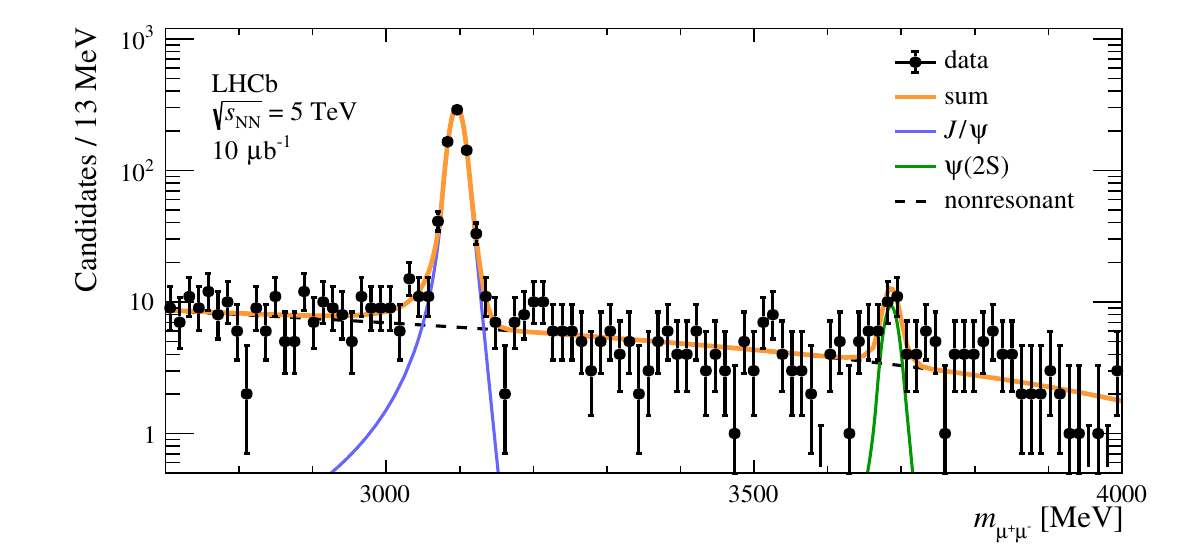}
  \end{center}
  \caption{
    The dimuon invariant mass spectrum in the range between $2.7$ and $4.0\gev$. The contribution of \jpsi (solid purple line) and 
    \psitwos (solid dark green line) mesons, and non-resonant background (dashed black line) are shown individually along with the sum of all contributions (solid orange line).  
    }
  \label{fig:mass}
\end{figure}

The yield of \jpsi mesons is estimated by fitting the dimuon invariant mass distribution to signal and background components. The
\jpsi and \psitwos mass shapes are modelled by double-sided Crystal Ball
functions~\cite{Skwarnicki:1986xj},
and the nonresonant background by an exponential function multiplied
by a first-order polynomial function.  The \psitwos parameters, aside from the
mean, are constrained to be the same as for the \jpsi meson.
The fit is performed in the range $2.7<m_\mumu<4\gev$. 
The dimuon mass distribution along with the fit projection is shown in Fig.~\ref{fig:mass}.

\begin{figure}[tb]
  \begin{center}
    \includegraphics[width=0.98\linewidth]{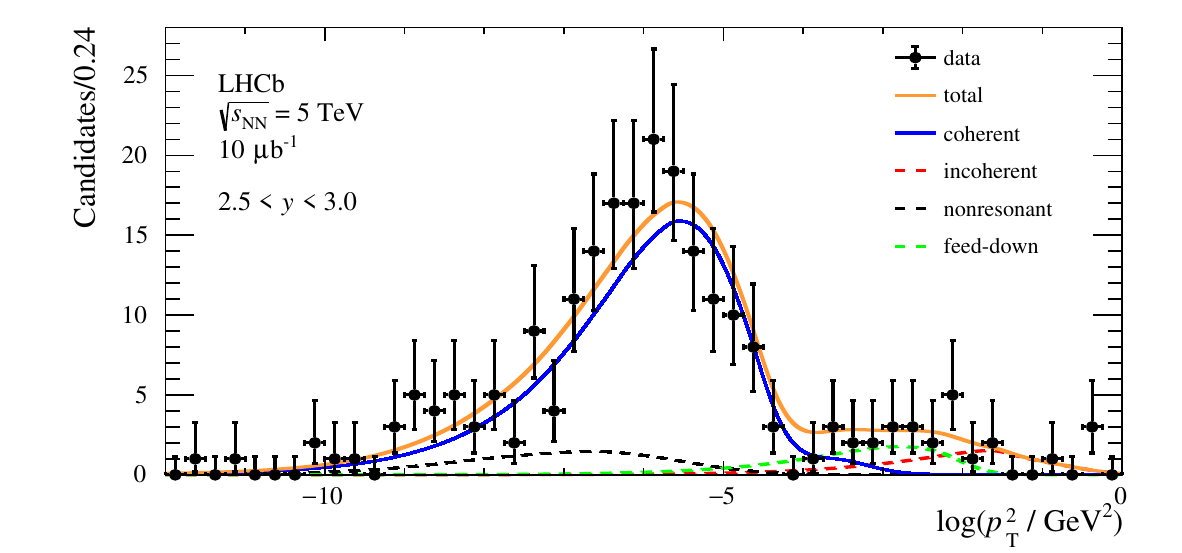}
    \vspace*{-0.5cm}
  \end{center}
  \caption{
    The  $\log(\ptjpsisq)$ distribution of dimuon candidates in the interval $2.5<y<3.0$, with \pt given in GeV, after all requirements have been applied.
     The solid orange line represents the combined fit to data; the solid blue line shows the coherent contribution; the incoherent component is displayed by the dashed red line; and the dashed green (black) line shows the feed-down (nonresonant) components.
     \update
  }

  \label{fig:fit}
\end{figure}

For the determination of the coherent yield two resonant background sources are considered: incoherent \jpsi photoproduction
and \jpsi meson feed-down from photoproduced \psitwos decays.  
In order to determine the signal yield in the presence of these
backgrounds, an unbinned maximum-likelihood fit to the natural logarithm of the transverse momentum squared, $\log(\ptjpsisq)$, of \jpsi candidates inside the chosen mass window is performed.
The signal and background probability density functions are estimated using the {\sc STARlight} generator\cite{starlight} and the LHCb detector simulation.
The amount of nonresonant background is constrained by the dimuon invariant mass fit. 
The feed-down background is assumed to have the same $\log(\ptjpsisq)$
distribution as simulated \decay{\psitwos}{\jpsi\pip\pim} decays, where the \jpsi is reconstructed and both pions escape the rejection requirements on additional tracks.
Figure~\ref{fig:fit} shows the $\log(\ptjpsisq)$ data distribution along with the fit projection in the rapidity interval $2.5<y<3$.
All  \jpsi yields are reported in Table~\ref{tab:yields}.
\begin{table}[]
    \centering
    \caption{Total and coherent \jpsi yields after the invariant mass and the transverse momentum fits, in \jpsi rapidity intervals. }
    \begin{tabular}{ccc}
    Rapidity $y$ & Total \jpsi yield & Coherent \jpsi yield \\\hline
%
\ensuremath{2.0-2.5}&\ensuremath{ \phantom{0}69\pm\phantom{0}9}	&\ensuremath{ \phantom{0}53\pm\phantom{0}8}	\\
\ensuremath{2.5-3.0}&\ensuremath{ 208\pm15}	&\ensuremath{ 153\pm14}	\\
\ensuremath{3.0-3.5}&\ensuremath{ 233\pm16}	&\ensuremath{ 176\pm15}	\\
\ensuremath{3.5-4.0}&\ensuremath{ 131\pm12}	&\ensuremath{ \phantom{0}95\pm11}	\\
\ensuremath{4.0-4.5}&\ensuremath{ \phantom{0}32\pm\phantom{0}6}	&\ensuremath{ \phantom{0}12\pm\phantom{0}5}	\\\hline

    \end{tabular}
    \label{tab:yields}
\end{table}

\subsection{Efficiency determination}

For any given \jpsi rapidity interval, the total efficiency is evaluated as the product of the  acceptance and the reconstruction and selection efficiencies. 
The acceptance includes the requirements on the kinematic properties of the \jpsi decay products,
and is evaluated using a sample of coherently photoproduced  $\jpsi\to\mumu$
events produced with the {\sc STARlight} event 
generator.
In this paper the charmonium polarisation is assumed to be the same as that of the photon and thus fully transverse as it is implemented in the STARlight event generator.
The reconstruction efficiency includes track reconstruction and muon identification.
The selection efficiency includes requirements on 
the SPD deposits,
VELO track multiplicities, and dimuon invariant mass. 
%
The hardware trigger efficiency is determined using simulated events,
calibrated with data. 
The software trigger efficiency
is measured 
using \jpsi candidates in events selected with a minimum bias trigger requiring at least one
VELO track. 
Partially reconstructed inclusive  \decay{\jpsi}{\mumu} candidates from proton-proton
collision data at a centre-of-mass energy $\sqs$=13\tev\ are used to evaluate the tracking efficiency~\cite{LHCb-DP-2012-003}.
The muon identification efficiency has been 
determined from simulated events generated with
the {\sc STARlight} generator in lead-lead collisions
and validated with lead-lead data.

The dimuon invariant mass requirement efficiency is determined using the
integral of the double-sided Crystal Ball function.
A similar method is used
to determine the efficiencies of the
multiplicity requirements on the 
VELO tracks and SPD deposits. 
The veto efficiency of the HeRSCheL detector activity is evaluated with a sample of nonresonant dimuon events from lead-lead data using a fit obtained from {\sc STARlight} simulated 
$\gamma\gamma\to\mumu$ events, and comparing the number of events 
satisfying and failing the requirement. The total efficiency 
for this selection is about 90\%.

\section{Systematic uncertainties}
\label{sec:sys}

Systematic uncertainties on the measured cross-section are 
considered from the integrated luminosity calculation,
the determination of the muon reconstruction and selection
efficiencies, the trigger efficiency,
the
mass fit signal model, the modelling of the feed-down background, and the knowledge of the $\jpsi\to\mu^+\mu^-$ branching fraction.
They are described below and summarised in Table~\ref{tab:sys}. The largest uncertainty originates from the integrated luminosity
determination and is estimated to be 13\%.

The uncertainties related to the \jpsi reconstruction
efficiency include effects on the
track reconstruction and muon identification, 
and they are dominated by the limited size of the 
control sample.
Several contributions to the selection 
efficiency are considered. 
The impact of the requirement on the SPD multiplicity 
is estimated from a data control sample. 
Effects
related to the dimuon invariant mass efficiency are taken from the uncertainty of the
integral of the double-sided Crystal Ball function.  
The inefficiency of the VELO track multiplicity requirement is found to
be negligible and no uncertainty is assigned.
The uncertainty due to the HeRSCheL selection is estimated by comparing the
efficiency evaluated in different samples, selected by applying
requirements that do not affect the signal. 

The efficiency of the hardware trigger is determined from simulated
events. It is compared to the efficiency obtained on a smaller data sample selected by independent triggers, and
the difference is taken as systematic uncertainty.
The software-stage trigger efficiency evaluation is cross-checked with an
independent estimation based on data, where events are selected with a
different trigger requirement. The efficiencies are consistent within statistical uncertainties, which are used to assign a systematic
uncertainty.

Possible cross-section variations associated to the signal model in the fit
to the dimuon invariant mass spectrum is assessed using an alternative model. 
A Bukin function~\cite{Bukin:2007} is used for the signal and the
difference in the signal yields with respect to the default fit is
assigned as uncertainty.

In order to estimate the systematic uncertainty due to the feed-down component, the \jpsi candidate selection is modified
to select a mixture of coherently and incoherently produced
\decay{\psitwos}{\jpsi\pip\pim} events.  After requiring the
reconstructed mass of the \psitwos candidates to be within $65\mev$ of
the known \psitwos mass~\cite{PDG2020}, 22 candidates are obtained.
In the simulation, \decay{\psitwos}{\jpsi\pip\pim} events are used as a proxy for all \decay{\psitwos}{\jpsi X} feed-down. The ratio between \decay{\psitwos}{\jpsi\pip\pim} candidates, where the pions left no VELO track,
and fully reconstructed candidates is determined and scaled assuming that \decay{\psitwos}{\jpsi\pip\pim} events are representative for all \decay{\psitwos}{\jpsi\PX} events.
From this ratio and the \jpsi yield, the number of \decay{\jpsi}{\mup\mun} candidates originating from \psitwos decays in the signal sample is determined to be $42.5\pm9.1$. 
This yield is assigned to the different intervals in meson rapidity using a template from simulated coherent \decay{\psitwos}{\jpsi\pip\pim} events.
The uncertainty is dominated by the limited size of the data sample.
\begin{table}[tb]
\caption{Systematic uncertainties considered for the differential
  cross-section measurement of coherent \jpsi production, relative to the central value.
  Uncertainty ranges correspond to variation over the rapidity intervals.
  The dominant uncertainty arises from the luminosity determination and is correlated over all intervals.
}
\begin{center}\begin{tabular}{lc}
\hline
Source  & Relative uncertainty (\%)      \\
\hline
Luminosity & $13.0$ \\
\jpsi reconstruction efficiency & $1.7$--$4.8$\\
Selection efficiency & $1.7$\\
HeRSCheL requirement efficiency & $1.0$\\
Hardware trigger efficiency & $1.0$\\
Software trigger efficiency & $1.0$\\
Mass fit model & $1.0-1.6$ \\
Feed-down background & $0.4-1.0$ \\
\jpsi branching fraction & $0.6$ \\
\hline
\end{tabular}\end{center}
\label{tab:sys}
\end{table}

\section{Results and discussion}
\label{sec:results}

\begin{figure}[tb]
  \begin{center}
    \includegraphics[width=0.78\linewidth]{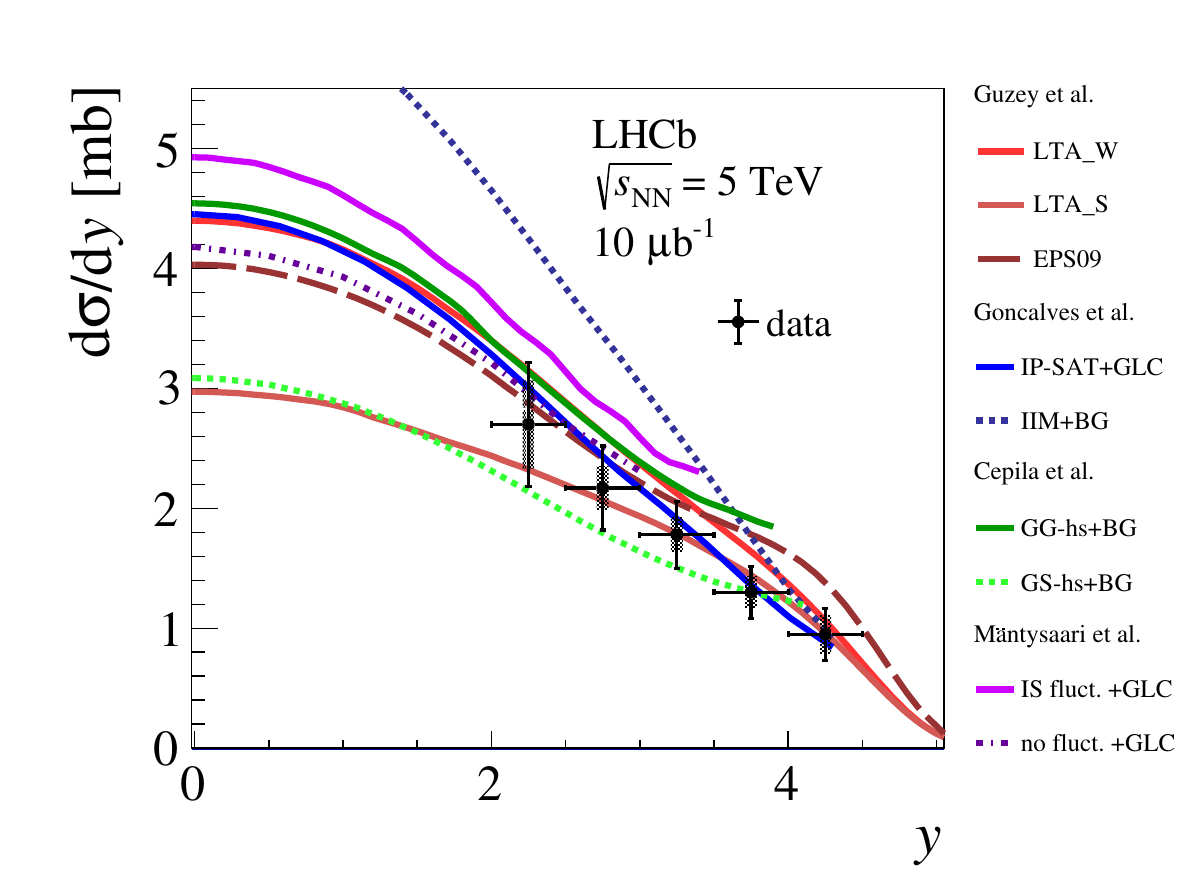}
   \vspace*{-0.5cm}
  \end{center}
  \caption{
  Differential cross-section  as a function of rapidity for coherent \jpsi production compared to 
  different phenomenological predictions~\cite{Guzey:2016piu,PhysRevC.84.011902,Goncalves:2017wgg,Cepila:2017nef,Mantysaari:2017dwh}.     
  The measurements are shown as points, where inner and outer error bars represent the statistical and the total uncertainties, respectively. 
  This includes the uncertainty on the luminosity and is therefore highly correlated.
  }
  \label{fig:xs}
\end{figure}

Using Eq.~(\ref{eqn:dsigmady}), the cross-section for coherent \jpsi production within the fiducial region is determined 
to be $$\sigma=\coherentcrosssectionuncsplit\mbarn,$$ where the first uncertainty is statistical, the second is systematic and the third is due to the luminosity determination.
The \jpsi candidates are reconstructed from dimuon final states, where the muons are detected within the pseudorapidity region $2.0 < \eta < 4.5$ and the \jpsi meson is required to have $\ptjpsi<1\gev$ and $2.0<\yjpsi<4.5$.
The coherent \jpsi production cross-section, measured in \jpsi rapidity intervals, is given in Table~\ref{tab:xs}. 
A comparison of the measured differential cross-section with the theoretical predictions discussed below is shown in Fig.~\ref{fig:xs}. 

In the model of Gon\c{c}alves et al.~\cite{PhysRevC.84.011902,Goncalves:2017wgg}, the cross-section is calculated within the framework of the Colour-Dipole model. Three different parametrisations of the dipole-proton cross-section (IIM, IP-SAT, bCGC), including saturation effects at low Bjorken-x, are combined with two different models of vector-meson wave functions, namely boosted Gaussian (BG) and Gauss-LC (GLC). All the parameters are tuned using HERA data\cite{Goncalves:2007sa,Aktas:2005xu,Chekanov:2002xi}. The solid (dashed) curves in Fig.~\ref{fig:xs} correspond to the IP-SAT+GLC (IIM+BG) model.
The combination of IIM~\cite{iancu2004saturation} with the boosted Gaussian wave function is disfavoured by the data.

The model from Cepila et al.~\cite{Cepila:2017nef} is a variation of the Colour-Dipole model. The main differences with respect to Gon\c{c}alves et al.~come from the parametrisation of the dipole-proton cross-section and the prescription used to propagate it to the dipole-nucleus scattering amplitudes. In this model, the Glauber-Gribov methodology (GG) or a geometric scaling between the nuclear saturation scale and the saturation scale in the proton (GS) are used. Both prescriptions are able to describe the data.

In the model proposed by M\"antysaari et al.~\cite{Mantysaari:2017dwh}, the cross-section is also calculated using the Colour-Dipole model including subnucleon scale fluctuations. Predictions with and without subnucleonic fluctuations using the IP-SAT parametrisation for the dipole-proton cross-section and the GLC for the vector-meson wave function are compared to this measurement.
Both prescriptions are able to describe the data.

The model provided by Guzey et al.~\cite{Guzey:2016piu} is based on a perturbative QCD calculation. The coherent \jpsi production cross-section on a proton target is calculated at leading order within the leading-log approximation.
Different models for the nuclear structure are used:
weaker (LTA\_W) and stronger (LTA\_S) nuclear shadowing scenarios with a leading twist nuclear shadowing model~\cite{Frankfurt:2011cs}  as well as EPS09~\cite{Eskola:2009uj} nuclear parton distribution functions. The measurement can be described by these models.

    \begin{table}[t]
          \caption{
        Measured cross-section $\sigma$ with breakdown of statistical, systematic and luminosity uncertainties measured as a function of the \jpsi rapidity. Note that the cross-sections are not normalised by the $y$ interval width. 
        }
      \label{tab:xs}
      \begin{center}
\begin{tabular}{ccccc}
      $y$ interval&$ \sigma\ [\mbarn]$&{Stat. $[\mbarn]$}&{Syst. $[\mbarn]$}&{Lumi. $[\mbarn]$}\\\hline
$2.0-4.5$ & 4.45 & 0.24 & 0.18 & 0.58 \\
\hline
$2.0-2.5$ & 1.35 & 0.19 & 0.06 & 0.17 \\
$2.5-3.0$ & 1.09 & 0.09 & $0.05$ & 0.14\\
$3.0-3.5$ & 0.89 & 0.07 & 0.04 & 0.12 \\
$3.5-4.0$ & 0.65 & 0.06 & $0.03$ & 0.08 \\
$4.0-4.5$ & 0.48 & 0.09 & 0.02 & 0.06 \\
\hline
\end{tabular}
      \end{center}
    \end{table}

The coherent photoproduction cross-section of \jpsi mesons in the forward region measured by the ALICE collaboration~\cite{Acharya:2019vlb} uses an integrated luminosity ten times larger in a smaller region of rapidity compared to this measurement.
In the kinematic region were the two experiments overlap the ALICE cross-section is reported about $35\%$ larger, with approximately a $30\%$ smaller uncertainty, leading to a discrepancy at the level of 2.3 standard deviations.

\section{Conclusions}
\label{sec:conclusion}

The coherent \jpsi production cross-section in lead-lead collisions at $\sqsnn=5\tev$, using a data sample collected by the LHCb experiment and corresponding to an integrated luminosity of about $\lumimeanrounded\invmub$, is measured to be $\coherentcrosssectionuncsplit\mbarn$, where the first uncertainty is statistical, the second is systematic and the third is due to the luminosity determination.
The measurement uses \jpsi mesons reconstructed in the dimuon final state with $\ptjpsi<1\gev$ and $2.0<\yjpsi<4.5$, where muons are detected within the pseudorapidity region $2.0<\eta<4.5$.
The cross-section is also measured in five \jpsi rapidity intervals and the results are compared to predictions from different phenomenological models.
Future measurements with different mesons and larger data samples will further constrain these models.

\section*{Acknowledgements}
%
%
\noindent We express our gratitude to our colleagues in the CERN
accelerator departments for the excellent performance of the LHC. We
thank the technical and administrative staff at the LHCb
institutes.
We acknowledge support from CERN and from the national agencies:
CAPES, CNPq, FAPERJ and FINEP (Brazil); 
MOST and NSFC (China); 
CNRS/IN2P3 (France); 
BMBF, DFG and MPG (Germany); 
INFN (Italy); 
NWO (Netherlands); 
MNiSW and NCN (Poland); 
MEN/IFA (Romania); 
MSHE (Russia); 
MICINN (Spain); 
SNSF and SER (Switzerland); 
NASU (Ukraine); 
STFC (United Kingdom); 
DOE NP and NSF (USA).
We acknowledge the computing resources that are provided by CERN, IN2P3
(France), KIT and DESY (Germany), INFN (Italy), SURF (Netherlands),
PIC (Spain), GridPP (United Kingdom), RRCKI and Yandex
LLC (Russia), CSCS (Switzerland), IFIN-HH (Romania), CBPF (Brazil),
PL-GRID (Poland) and NERSC (USA).
We are indebted to the communities behind the multiple open-source
software packages on which we depend.
Individual groups or members have received support from
ARC and ARDC (Australia);
AvH Foundation (Germany);
EPLANET, Marie Sk\l{}odowska-Curie Actions and ERC (European Union);
A*MIDEX, ANR, Labex P2IO and OCEVU, and R\'{e}gion Auvergne-Rh\^{o}ne-Alpes (France);
Key Research Program of Frontier Sciences of CAS, CAS PIFI, CAS CCEPP, 
Fundamental Research Funds for the Central Universities, 
and Sci. \& Tech. Program of Guangzhou (China);
RFBR, RSF and Yandex LLC (Russia);
GVA, XuntaGal and GENCAT (Spain);
the Leverhulme Trust, the Royal Society
 and UKRI (United Kingdom).


\addcontentsline{toc}{section}{References}
\setboolean{inbibliography}{true}
\bibliographystyle{LHCb}
\bibliography{main,LHCb-PAPER,LHCb-CONF,LHCb-DP,LHCb-TDR,local}
 
\newpage

\centerline
{\large\bf LHCb collaboration}
\begin
{flushleft}
\small
R.~Aaij$^{32}$,
C.~Abell{\'a}n~Beteta$^{50}$,
T.~Ackernley$^{60}$,
B.~Adeva$^{46}$,
M.~Adinolfi$^{54}$,
H.~Afsharnia$^{9}$,
C.A.~Aidala$^{86}$,
S.~Aiola$^{25}$,
Z.~Ajaltouni$^{9}$,
S.~Akar$^{65}$
% ...

{\footnotesize \it

% ...
$^{9}$Universit{\'e} Clermont Auvergne, CNRS/IN2P3, LPC, Clermont-Ferrand, France\\
% ...
$^{25}$INFN Sezione di Milano, Milano, Italy\\
% ...
$^{32}$Nikhef National Institute for Subatomic Physics, Amsterdam, Netherlands\\
% ...
$^{46}$Instituto Galego de F{\'\i}sica de Altas Enerx{\'\i}as (IGFAE), Universidade de Santiago de Compostela, Santiago de Compostela, Spain\\
% ...
$^{50}$Physik-Institut, Universit{\"a}t Z{\"u}rich, Z{\"u}rich, Switzerland\\
% ...
$^{54}$H.H. Wills Physics Laboratory, University of Bristol, Bristol, United Kingdom\\
% ...
$^{60}$Oliver Lodge Laboratory, University of Liverpool, Liverpool, United Kingdom\\
% ...
$^{65}$University of Cincinnati, Cincinnati, OH, United States\\
% ...
$^{68}$Syracuse University, Syracuse, NY, United States\\
% ...
$^{86}$University of Michigan, Ann Arbor, United States, associated to $^{68}$\\
\bigskip
% ...
\medskip
}
\end{flushleft}

\end{document}